%% file: user_session.tex
\documentclass[sigconf]{acmart}
\usepackage{tabu}
\usepackage{subcaption}
\usepackage{booktabs} % For formal tables
\newcommand{\xhdr}[1]{\vspace{1.7mm}\noindent{{\bf #1.}}}

\setcopyright{none}
\acmDOI{00000}

\acmISBN{00000}
\acmYear{2017}
\copyrightyear{2017}
\acmPrice{15}

\begin{document}
\title{
Categorizing User Sessions at Pinterest
}

\author{Dorna Bandari}
\affiliation{%
  \institution{Pinterest }
}
\email{dorna@pinterest.com}

\author{Shuo Xiang}
\affiliation{%
  \institution{Pinterest}
}
\email{sxiang@pinterest.com}

\author{Jure Leskovec}
\affiliation{%
  \institution{Pinterest}
}
\email{jure@pinteres t.com}

\begin{abstract}
Different users can use a given Internet application in many different ways. The ability to record detailed event logs of user in-application activity allows us to discover ways in which the application is being used. This enables personalization and also leads to important insights with actionable business and product outcomes.

Here we study the problem of user session categorization, where the goal is to automatically discover categories/classes of user in-session behavior using event logs, and then consistently categorize each user session into the discovered classes. We develop a three stage approach which uses clustering to discover categories of sessions, then builds classifiers to classify new sessions into the discovered categories, and finally performs daily classification in a distributed pipeline. An important innovation of our approach is selecting a set of events as long-tail features, and replacing them with a new feature that is less sensitive to product experimentation and logging changes. This allows for robust and stable identification of session types even though the underlying application is constantly changing. We deploy the approach to Pinterest and demonstrate its effectiveness. We discover insights that have consequences for product monetization, growth, and design. Our solution classifies millions of user sessions daily and leads to actionable insights.

\end{abstract}

 
\maketitle
\input{main}

\bibliographystyle{unsrt}
\bibliographystyle{abbrv}

\bibliography{sigproc} 

\end{document}

%% file: main.tex
\section{Introduction}
  %This work examines the problem of finding session-wise use cases of an Internet application, using user event logs. In other words, we find the different ways people use the app in a single session, and the relative scale of each use-case. Additionally, we propose a practical method for implementing our solution in large scale web applications.

%perform a variety of analysis tasks such as user segmentation, page classification, as well as predictive tasks such as collaborative recommendations
%  \item \textbf{What is the problem?} REWORD
 
%Internet applications can record a constant stream of event logs as the user interacts with their application. 
%With recent advancements in distributed computing and storage, 
As users interact with Internet applications a constant stream of event logs gets recorded. The ability to collect detailed event logs of user behavior presents the opportunity to gain important insights into the usage patterns of the applications, and also enables product personalization and recommendations. Particularly significant is the problem of user session categorization, where the aim is to classify types of sessions based on usage patterns.

%in order to map those session types to different user intent in a given application.

%why did the user use the website or application in a specific visit. 
%This has been studied in an area of research called Web Usage Mining. 
%the ability to collect minute details of user behavior presents great opportunities as well as challenges. Given the scale and diversity of possible actions in these applications,

%While collecting minute details of user behavior presents the potential 

%  \item \textbf{Why is it important?} 
%Unlocking the hidden insights in these logs can lead to advancements in many areas from product personalization to business strategy. 

User session categorization is an important problem as it leads to critical and actionable insights about the application. For example, discovering that a user commonly uses the application for casual browsing versus searching enables creation of a personalized home page, with buttons or links that make navigation easier. Additionally, using the information about users' previous behavior, applications can personalize recommendations, search ranking, as well as advertising.
Furthermore, categorizing sessions has significant value for business insights and strategy. Discovering session categories of an application and their relative scale leads to insights, such as estimating the strategic and monetary value of different use cases of the application, tracking changes in user behavior over time, and analyzing effects of product experiments. Such insights can guide the company in setting the right business and product goals as well as deciding which metrics to track and optimize. 

%  \item \textbf{Why is it hard?} 
% A Pinterest user typically visits several pages in the application and performs a combination of actions in each, and potentially multiple times. There are often cycles in the path of user as they navigate the application. 
 
There exists two main challenges in categorizing user sessions. First is the large scale and high dimensionality of the underlying data. User sessions often contain thousands of actions in a single session, such as clicking on various links, scrolling, searching, etc. Many of these actions may be unique. Moreover, an application may have millions of unique new sessions every day \cite{Kou2012}. This makes discovering and categorizing patterns in user behavior extremely difficult, especially in applications with diverse and complex use cases (e.g. social networks, content discovery platforms, messaging platforms, and lifestyle applications). The second challenge is that user logs tend to be a constantly changing and unreliable data source \cite{Katal2013}. As new product experiments are rolled out and the user interface changes, logged events tend to change. Therefore it is challenging to build a model on this dataset that will remain stable over time through most experiments and logging changes. 

% why hasn't it been solved before
%The field of \emph{web usage mining} studies how to categorize user in-session behavior . 
The problem of analyzing user sessions using event logs and clickstream data in online computing applications has been studied in the field of \emph{web usage mining}~\cite{Mobasher:2000,Eirinaki:2003,Yan:1996,Cho2002329,srivastava2000web}.
Popular solutions in web usage mining include association rule learning and sequential pattern mining. However, these approaches were proposed for finding patterns and rules, while we focus on broad categorization of user sessions. Additionally, the majority of the solutions are too computationally complex to be deployed to large-scale Internet applications. Moreover, these methods generally do not address the problem of stability in the face of logging changes and product experimentation.
%either too computationally intensive for anything but the most basic applications, or involve simplifying assumptions that result in too much loss of information. One potential approach is using sequential pattern mining techniques

  %Over time, new events are added and some older events disappear as UI elements are moved around in the pages
  %Therefore these logs can be extremely difficult to decipher in applications with diverse and complex use cases (i.e. social networks, messaging platforms and lifestyle applications), .
  %would enable analysis without losing important information is a difficult problem.
  %and the problem of handling logging changes as a result of experimentation and user interface changes. 
% \item \textbf{Why hasn't it been solved before?}
% References, more critical, many citations, more precise ...
 
  %Popular approaches are either too computationally intensive for anything but the most basic applications, or involve simplifying assumptions that result in too much loss of information. One potential approach is using sequential pattern mining techniques. Computational complexity of this method is prohibitive in applications where the use cases involve a long sequence of actions. Another approach is to find association rules, i.e. the actions that tend to be present in the same sessions. This results in too much granularity and does not lead to finding broad latent use cases or tasks in sessions.
  
%  \item \textbf{What are the key components of my approach?} 

\xhdr{Present work}
Here we develop a three-step approach for classifying usage sessions based on user event logs. Our solution has been deployed at Pinterest, a content sharing platform with over 150 million monthly users \cite{silbermann2016}, and classifies tens of millions of user sessions per day. Additionally, we have evaluated our methodology on over six months of data and our method achieves stability in the face of logging changes and product experimentation. The three steps to our solution are as follows: 
  \begin{itemize}
  \item We cluster a sample of sessions and create a labeled dataset for this sample. 
  \item We use this labeled data to build a predictive model that is designed to achieve stability over time.
  % BK COMMENT: Is it really online? Isn't it an offline process?
  \item We deploy the model in a distributed computational pipeline that performs daily scoring of all Pinterest user sessions.
  \end{itemize}

  %Our clustering method borrows ideas from the field of NLP\cite{Salton:1989:ATP:77013}, which
  
  % IN INTRO AS PART OF NARRATIVE - DEF PUT
%Our contributions are 1. We use TF-IDF weights for the actions, which is an important factor when the number of events are high and some of the actions are very prevalent across sessions (such as scrolling or clicking), 2. 

Our methodology offers the following benefits: Our clustering method decreases the complexity of the problem without loss of important information. Instead of considering the session as a sequence of events in time, we consider each user session as a document, and each user action as a word in the document. We then apply document clustering methods to user sessions, which results in stable clusters that reveal major per-session use cases of Pinterest. Our predictive model focuses on the most robust and fundamental user events on the application, and it is not affected by logging changes and product experimentation. For example, logging a temporary new event on the application will not change the results of our daily classification. This is done by dividing user events into two groups, \emph{Scoring features} and \emph{Long tail features}, where Scoring features include only fundamental actions on the application. We define a new feature called  \emph{Noise feature} that replaces the set of Long tail features, decreasing sensitivity of our model to changes in them. Our classifier achieves more than 85\% prediction accuracy.

% \item \textbf{Results?}
%\textbf{(Probably can't say some of this)} 
%Give numbers, so it'smmore real. X number of sessions over X amount of time

 After deploying the models on Pinterest, we identified six major session categories, each corresponding to a distinct and interpretable use case. We analyzed a sample of user sessions in the month of July 2016 (8.2 million sessions), and present a selection of insights. Our analysis revealed important patterns in user behavior. We found that content consumed by users in different session categories varies significantly, with low-intent session categories involving content that are aspirational or merely entertaining, and high-intent session categories involving highly practical content. We found that newer cohorts on Pinterest differ from older cohorts in session categories they utilize. Additionally, advertisement revenue differs substantially in different session categories, identifying opportunities for increasing revenue. We also present the difference in length of sessions and transition probability of session categories.

% Our analysis reveals important patterns in user behavior. For example, content consumed in different session types varies significantly, with low-intent session types involving content that are aspirational or merely entertaining, and high-intent session types involving practical content. We learned that 

% * <dorna.bandari@gmail.com> 2017-02-15T21:39:16.835Z:
% 
% trying out commenting
% 
% ^ <dorna.bandari@gmail.com> 2017-02-15T21:39:27.466Z.
%  \item \textbf{Summary of Contributions} 
The rest of the paper is organized as follows. 
In Section \ref{sec:rel} we will list existing solutions for web usage mining. In Section \ref{sec:problemstat} we will describe the specific problem we aim to solve, and the associated dataset. In Section \ref{sec:overview} we will describe our proposed solution in detail, with Section \ref{sec:clustering} describing cluster discovery, Section \ref{sec:classifiers} detailing the design of the prediction models, and Section \ref{sec:deployment} describing our system in production. Section \ref{sec:experiments} lists the experiments and analyses that informed our design choices. In Section \ref{sec:deploymentinsights} we list specific details of our final deployed models, and illustrate some of the key insights we found from this work.

% --------------------------------------------------------

\section{Related Work}
\label{sec:rel}
Web usage mining or clickstream mining has been an active area of research~\cite{Agrawal:1993,Cooley}. Data mining algorithms have been applied to the user sessions for a variety of purposes, such as personalization of algorithms and applications \cite{Yan:1996,Eirinaki:2003}, recommendation systems \cite{Cho2002329}, as well as obtaining business intelligence and aiding strategy \cite{Tiwari}. Most of the research lies in one of two broad areas: association rule learning, which generally ignores the ordering between pages or actions, and path mining methods (including sequential pattern mining and clustering), which generally takes the order into account. 

Association rule learning involves algorithms that discover rules that define relationships between items in a set (e.g., web pages) \cite{Agrawal:1993,Liu}. An example would be discovering that users who visit a web page $A$ and take a specific action on that page are most likely to also visit web page $B$. Commonly the order of pages or actions is ignored. This class of algorithms have a variety of practical use cases, such as market and risk management and web personalization \cite{Shaw,Mobasher}. The drawback of association rule mining methods is that the rules will not result in broad general categorization of sessions, while our method is used to find major user behavior categories.

Sequential pattern mining methods are the class of algorithms that discover patterns in a sequence of items \cite{Agrawal:1995}, which in the case of web mining could be clicks or pages that users navigate to. The patterns may be found based on frequency of occurrence or some other measure of importance of a sequence. These methods are mainly practical for simple applications or parts of an application, such as a purchase funnel in an e-commerce application where the possible paths or actions are limited. Our method is designed specifically for large scale applications with complex navigation paths and cycles.

Clustering methods involve creating a set of features from the path a user takes as they navigate the web site. Shahabi et al. use time spent on each page as the main feature for the clustering \cite{Shahabi}. Fu et al. categorize pages on a web site using an attribute-oriented approach \cite{Han:1992} in order to decrease the dimensionality prior to clustering the sessions using hierarchical clustering \cite{Fu}. Heer et al. use a combination of features of the pages, such as TF-IDF \cite{Salton:1986} of the content, as well as the path in order to find session clusters \cite{Heer02miningthe}. These methods were proposed for use on web site pages and links, rather than minute user actions, therefore their computational complexity would make their use for clickstream data in the scale of modern applications prohibitively large. Our method on the other hand is built for applications with millions of daily user sessions.

Jin et al. \cite{Jin:2004} propose using Probabilistic Latent Semantic Analysis (PLSA) to analyze web usage, and Xu et al. use PLSA to group web pages \cite{plsa}. These methods are closest to our work, in that they reduce the dimensionality of the logging data by creating a matrix of sessions by unique web pages (or click actions). In \cite{Jin:2004}, the matrix consists of only binary values, while in \cite{plsa}, time spent per web page is  used to create the data. Neither of these methods address the issues of instability of the results in face of product experimentation and logging changes, while our system involves practical solutions to address this common problem.

\section{Problem Statement}
\label{sec:problemstat}
When a user interacts with an Internet application, their actions are logged in the order they were performed. The aim then is (1) to discover types/classes of user behavior using event logs, and (2) to consistently categorize all usage sessions into the discovered classes. The focus of this paper is behavior in a single usage session so that categorization of all sessions can be performed on a daily basis.

There are several unique challenges and considerations when building a production system for session categorization. First is that modern Internet applications are complex and are available on multiple platforms. For example, Pinterest is available on six different platforms (iPhone, iPad, Android mobile, Android tablet, web, and mobile Web) with each having distinct user interface elements and some distinct user actions. 

% While our deployed system has a different set of models for each application, we only refer to one set of models as we describe the methodology.
%
%Note that we repeat this process for each device and have different models for each, because event types are not the same across devices. Additionally usage patterns are very different across devices, which leads to different session types in each device. In the following sections 

Second, on Pinterest, like most complex Internet applications, there are several main activities a user can take part in. Users can view as well as save pieces of content which we call \emph{pins}. They can click on a pin to visit the original web site that content links to, and they can search Pinterest for specific content. They can message each other, comment on pins, and read their notifications. In addition to these major activities, there are hundreds of other possible minor actions and user interface interactions a user can engage in. The challenge here is that there is no \emph{a priori} clear way of knowing which of these actions are important to determine usage types and which ones can be discarded as noise. 

\section{Methodology}
\label{sec:overview}

\begin{table}[t!]
  \caption{Example of four sessions and the list of events associated with each session}
  \label{tab:exsess}
  \begin{tabular}{c m{7cm} }
    \toprule
   ID & User events in session  \\
    \midrule
    1 & {SEARCH, SEARCH, CLICKTHROUGH, CLICK, CLICK, SCROLL, SCROLL, SEARCH, VIEW, SEARCH}\\
    2& {CLICK, PIN\_VIEW, TAP, SEND\_MESSAGE, READ\_MESSAGE, SEND\_MESSAGE }\\
    3& {PIN\_VIEW, REPIN, REPIN, SCROLL, TAP, REPIN, SCROLL, SCROLL}\\
    4& {PROFILE\_VIEW, PIN\_VIEW, PROFILE\_VIEW}\\
  \bottomrule
\end{tabular}
\end{table}

\begin{figure}[t]
  \centering
  \includegraphics[width=7cm,keepaspectratio=true,trim={4cm 7cm 3cm 7cm}]{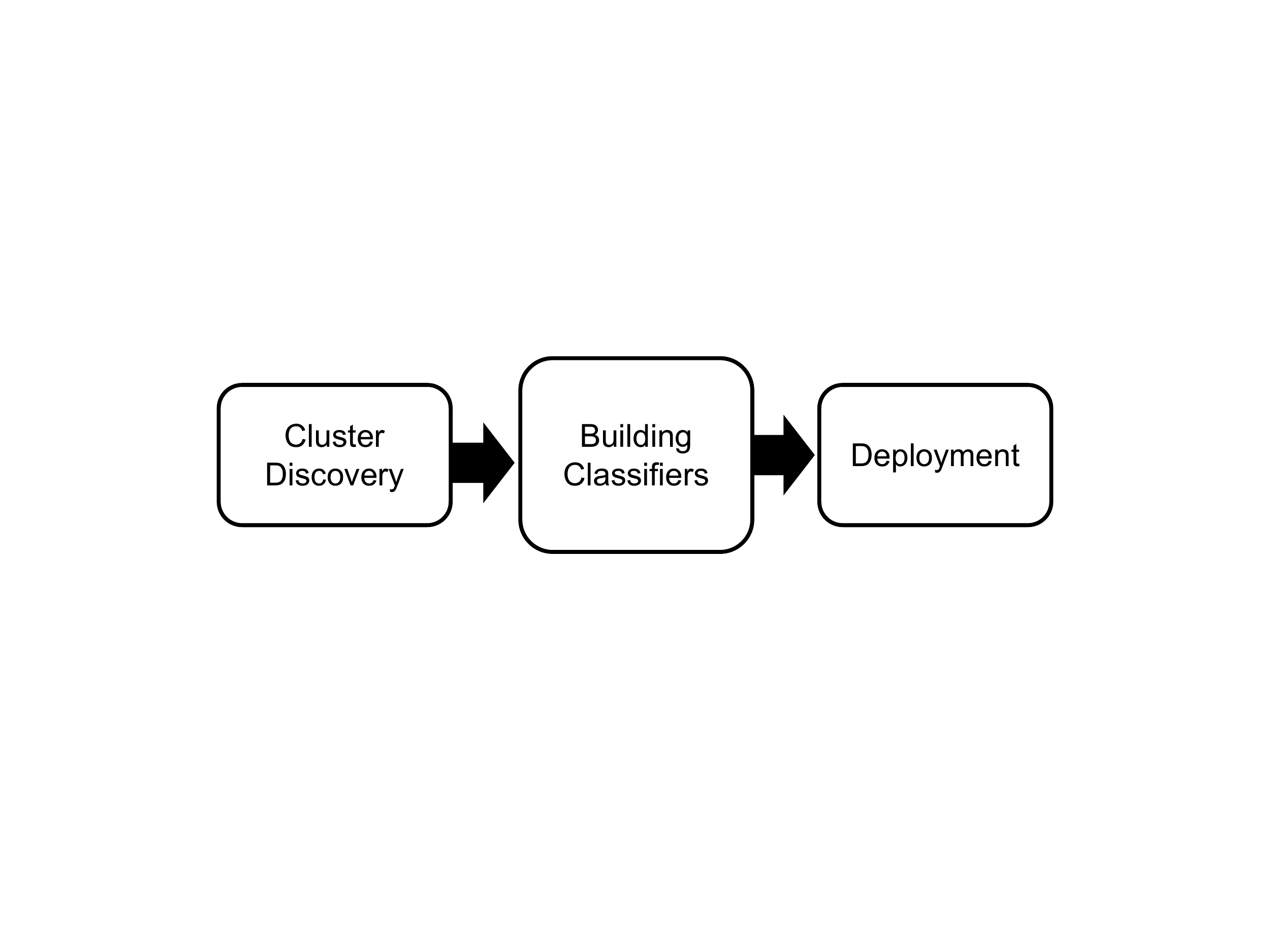}
  \caption{System overview}
  \label{fig:overview}
\end{figure}

Our proposed solution consists of three stages (Figure \ref{fig:overview}). First, we discover session clusters using a sample of daily sessions. Next, we build a classifier that classifies each session into one of the clusters. Finally, we deploy the model in a distributed computational pipeline. 

\subsection{Data Preprocessing}
First we process the raw event logs in order to assign each event to the unique user who performed the action, clean the data to remove spam and bot data, and filter out background event logs (i.e. events that are not triggered explicitly by user action, but by our application), and remove sessions that are too short  to have had a purpose. A discussion of data preprocessing steps is beyond the scope of this paper.

We then sessionize the logs. There are a few different ways to define a user session \cite{Spiliopoulou:2003}.  In this work we take a time-lapse approach to defining user sessions, meaning that a user session is defined as a group of user actions that occur in an isolated period of time, with a pre-defined gap of inactivity before and after. This gap is found using the distribution of inter-action times on each device. 

The cleaned, preprocessed data will have the form given in Table \ref{tab:exsess}, with each row corresponding to a specific user session.

\subsection{Cluster Discovery}
\label{sec:clustering}

In order to discover session-wise use cases we propose to cluster sessions based on user actions. We consider the list of user actions in a session as a document, with each user action as a word in the document. 
We use a sample of sessions in one day, and filter the actions to remove the ones that occur in fewer than 5\% of sessions. We then find the TF-IDF of the actions \cite{Salton:1986}. In TF-IDF representation, the number of times an action occurs in a session is normalized by the inverse document frequency (IDF). IDF reduces the weight of more frequent actions in sessions. This means common actions such as \emph{CLICK} or \emph{SCROLL} in a mobile session would have lower weight in the clustering.

We normalize the vector of weights for each user session. The purpose of the normalization is to characterize each session by the actions that are dominant in that session, irrespective of session length.

We then use principal component analysis (PCA) to reduce the dimensions of the data, and cluster the projected results using K-Medoids \cite{kaufman1987clustering} algorithm. Experiment results are listed in Section \ref{sec:evalclust}.

Summary of the cluster discovery method is as follows:
\begin{enumerate}
    \item Sample user sessions in one day.
    \item Filter out events that occur in fewer than 5\% of sessions.
    \item Find the TF-IDF weights for user events in every session.
    \item Normalize the vector in each session. 
    \item Reduce the dimensions using PCA.  
    \item Cluster the projected results using K-Medoids. 
\end{enumerate}

\subsection{Building Classifiers}
\label{sec:classifiers}

The next step in our method is to build a multiclass classifier that will be deployed in production in order to assign session categories to daily user sessions. Note that session categories can be defined such that they map exactly to each session cluster found in Section \ref{sec:clustering}, or alternatively, some clusters can be combined together to define a major session category. In our deployed system, after qualitative assessment of session clusters, we combined the clusters into 7 major session categories that correspond to major use cases of Pinterest. This simplification also made the categories consistent across different Pinterest applications.
 
One of the main reasons for building classifiers is achieving stability over time in spite of product experimentation and changes in user interface. Our classifier overcomes this issue by using a smaller subset of events that tend to be most stable in time.

\xhdr{Scoring features versus long tail features} 
User events are not equally robust over time. For example, events associated with minute user interactions with user interface elements may vary with minor design changes. E.g. changing the size of some elements on the page may change the frequency of triggering of another event. These changes may not be the result of fundamental changes in user behavior. The stability issue is caused by the following facts:

\begin {itemize}
  \item Minor user interface changes can change the proportion of logged events, possibly leading to new erroneous clusters. 
  \item When a new feature is being experimented on, it affects a small subset of users. This leads to artificially high IDF for user events associated with the new feature, as they seem rare. This could lead to new false clusters.
\end{itemize}

We aim to retrain the model once every three months. Consequently, we must choose a subset of events that would remain relatively steady in a three month period. However, note that we cannot assume stability of events that were not involved in past experiments and logging changes, since they may be affected by future ones. For this reason, it is best to only select fundamental actions on the application using prior knowledge, and only select additional events if prediction accuracy is not satisfactory. In other words, we aim to choose as few events as possible, while having acceptable prediction accuracy for every session category. 

Another method for selecting the scoring features is by analyzing the historical daily counts of events. In this case, we would detrend and deseasonalize the daily counts of each event over a three month period, and find the variance of errors. Then, we would select events in the decreasing order of variance, stopping when an acceptable prediction accuracy occurs.

We name the chosen, stable events the \emph{Scoring features}, and the rest the \emph{Long tail features}.

\xhdr{Noise feature} Since we remove the long tail events from the prediction, we lose information in long tail user behavior, which we have labeled the \emph{Noise} category. Therefore we have very poor accuracy in the \emph{Noise} category, as demonstrated in Table \ref{tab:classifiererrors}. In order to improve the accuracy in this class, we will create a new feature called \emph{Noise feature}. It is defined as
\begin{equation}
\label{eq:ltfeature}
\eta_i = \frac{ \displaystyle\sum_{f \in F_s } w_i(f)}{\displaystyle\sum_{f \in \{F_s \cup F_{lt}\} }  w_i(f)}.
\end{equation}

\(F_s\) is the set of scoring features, \(F_{lt}\) is the set of long tail features, and \(w_i(f)\) is the TF-IDF weight of feature \(f\) in session \(i\). Since long tail events are possibly changing drastically over time, we normalize the \(\eta_i\) feature by the mean of this feature over all sessions. This step is important in ensuring that a sudden change in weight of some of the long tail features will not create large shifts in size of session categories.

Finally, using the Scoring features along with the Noise feature, we build a random forest model to predict the session categories. Prediction accuracy of the models trained with and without the new Noise feature is given in Table \ref{tab:classifiererrors} in Section \ref{sec:evalclass}.

Summary of the classification step is given below.

\begin{enumerate}
    \item Start with the labeled dataset from clustering method, a sample of user sessions with session category labels. 
    \item Find TF-IDF of all events. 
    \item Create the Noise feature using Equation \ref{eq:ltfeature}, and normalize it across all sessions. 
    \item Normalize the vector of scoring features in each session. 
    \item Train a random forest model using data from step 2 and 3.
\end{enumerate}

\subsection{Deployment}
\label{sec:deployment}

Having considered the downstream workflows as well as Pinterest's own computation infrastructure, we decide to use two separate pipelines to accommodate computational tasks that are of various deployment frequencies. Specifically, we have a daily pipeline that performs the aforementioned data preprocessing and scoring, since the results of both tasks will be consumed daily by downstream jobs. On the other hand, cluster discovery and model update are currently carried out on demand. Though it is possible to combine these tasks into a single workflow, we believe the current setup provides better balance between development velocity and scalability.

\xhdr{Daily scoring} Once a model is obtained, it is deployed online to compute scores for each session in our entire context session logs. Due to the large amount of daily session data, a distributed computational pipeline is necessary and we implement an end-to-end scoring pipeline in Apache Spark. We use Pinball\footnote{ https://github.com/pinterest/pinball} to compose a workflow that will be scheduled daily.

\xhdr{IDF}
Rarity of events may change as product experiments are rolled out to different proportion of user base. This means that even user sessions that are not part of the experiment would be affected by the experiment, since the IDF is calculated over the entire data set. So while in this case we cannot do much about the user sessions in the experiment, we should ensure that session not in the experiment will not be affected. Therefore, given that the prediction model will take TF-IDF weights of the subset of events, the IDF should not be part of the computation pipeline, and it should only be updated every time the model is re-trained. 

\xhdr{Updating the model} 
Model update usually comes with recomputing or back-filling existing data. To better support retrospective experimentation, we have stored versioned data on a persistent storage system and by default show the latest result. 

\section{Experiments}
\label{sec:experiments}
\begin{figure}
  \centering
  \includegraphics[width=7cm,keepaspectratio=true,trim={0cm 0cm 0cm 0cm}]{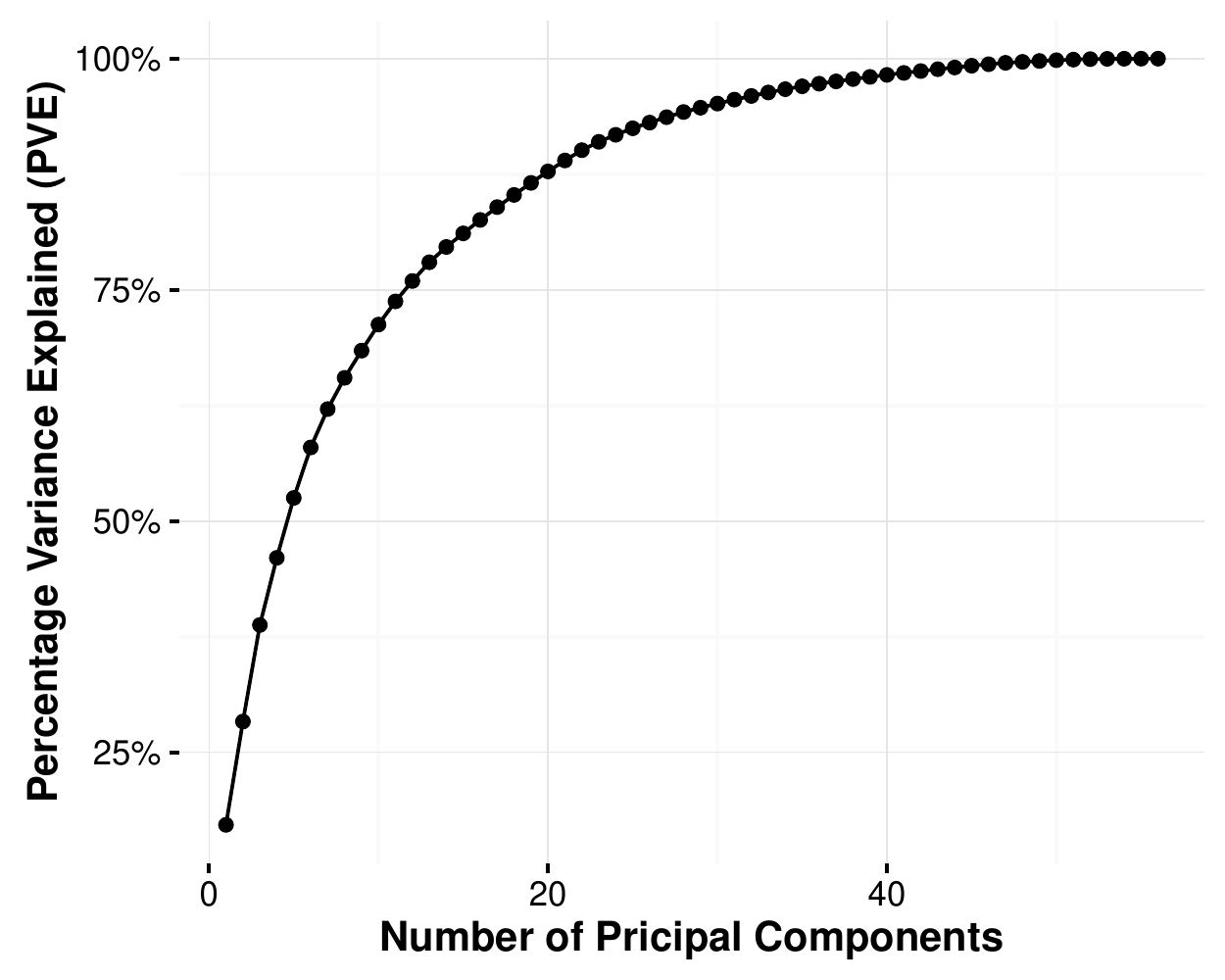}
  \caption{On iPhone devices we selected 14 principal components, which explains 80\% of variance in the data.}
  \label{fig:pca}
\end{figure}

\begin{table}
            \caption{Clustering trials on the iPhone Pinterest application. We chose 13 clusters, since clusters are found to be stable, and Noise cluster is smaller than the case with 8 clusters.}
  \label{tab:expclust}
  \begin{tabular}{l c c c }
    \toprule
 Number of clusters & 18 &	13&	8  \\
    \midrule
\% in Noise cluster  & 9\% & 15\% & 21\% \\
Avg. Stability, Jaccard distance  & 0.91 & 0.89 & 0.96 \\
Avg. Silhouette  & 0.3 & 0.33 & 0.39 \\
\% Clustered with stability of +0.7   & 96\% & 100\% & 100\% \\
  \bottomrule
\end{tabular}
\end{table}

\begin{table*}[t!]
  \caption{Confusion matrix for final classifier for Pinterest iPhone application, on the test dataset. The session categories are named using qualitative assessment of the data. }
  \label{tab:confusion}
  \begin{tabular}{l c c c c c c c }
    \toprule
Category &	Browse &	Clickthrough	&Notification &	Noise  &Retrieval &	Repin &	Search\\
\midrule
Browse  &93.9\%	&0.9\%	&0.2\%	&2.5\%	&0.7\%	&1.0\%&	0.7\% \\
Clickthrough & 0.8\%	&94.2\%	&0.1\%	&2.1\%&	1.2\%&	0.6\%&	1.1\%\\
Notification   & 0.4\%	&0.1\%	&97.2\%	&1.5\%&	0.5\%&	0.2\%&	0.0\%\\
Noise   &8.1\%	&2.2\%	&1.1\%	&73.2\%&	9.5\%&	3.9\%&	2.0\%\\
Retrieval   &2.6\%	&2.1\%	&0.3\%	&9.5\%&	82.0\%&	1.8\%&	1.8\%\\
Repin   & 2.1\%	&0.4\%	&0.2\%	&3.4\%&	0.9\%&	92.0\%&	1.1\%\\
Search   & 0.6\%	&0.3\%	&0.1\%	&0.8\%&	0.3\%&	0.8\%&	97.2\%\\
  \bottomrule
\end{tabular}
\end{table*}

\begin{table}[t!]
  \caption{Misclassification error rate in each class, when classifier uses only the subset of robust events, the subset of robust events with the Noise feature, and all events. The middle column has much lower error in the Noise category, while achieving robustness over time by eliminate the more volatile features.
  }
  \label{tab:classifiererrors}
  \begin{tabular}{l m{1cm} m{2cm} m{2cm} }
    \toprule
Cluster  &  
Scoring \newline  features \newline  only & 
Scoring \newline features +\newline Noise  feature & 
Scoring +\newline Long-tail  features \\
\midrule
Browse   & 6\% & 6\% & \%3 \\
Clickthrough   & 6\% & 6\% & \%2\\
Notification   & 3\% & 3\% & \%1\\
\textbf{Noise}   & \textbf{80\%} & \textbf{27\%}  & \textbf{\%7}\\
Retrieval   & 20\% & 18\%  & \%5\\
Repin   & 7\% & 8\%  & \%5\\
Search   & 2\% & 3\%  &  \%2\\
  \bottomrule
\end{tabular}
\end{table}
In this section we present the results of some of the experiments that lead to some of our important design choices. Note that some of the choices were made through qualitative assessment of the results by the authors and our internal partners. This is an important step in implementing such a system for practical applications.

\subsection{Evaluating Clustering} 
\label{sec:evalclust}
We select the number of principal components such that more than 80\% of variance is explained by the projected features. Figure \ref{fig:pca} illustrates the Percentage Variance Explained (PVE) plot for the iPhone Pinterest application. The elbow in this plot occurs at around 80\% value, therefore we choose 14 principal components in this case.

In order to validate the clustering method and select the right number of clusters, we considered a few different factors. One is qualitative assessment and usability of clusters by our internal partners. Our internal conversations lead us to choose broad types of clusters as opposed to smaller clusters. Additionally, we considered the percentage of sessions that were clustered in the \emph{Noise cluster}, with the aim of decreasing the size of this cluster. Other metrics we assessed are Silhouette score \cite{Rousseeuw:1987:SGA:38768.38772}, which measures the gap between clusters, and cluster stability \cite{Lange,Hennig07cluster-wiseassessment}, which measures reproducibility of clustering results. For cluster stability, we created 50 randomly selected subsets of the data set, each having half of the original data, as suggested in \cite{Hennig07cluster-wiseassessment}. We found the average Jaccard similarity of each original cluster membership to the one found in each subset. We assumed a cluster is stable if more than 70\% of the points were assigned to the original cluster. 

  Table \ref{tab:expclust} summarizes three of the clustering options we considered, along with the metrics noted above. We chose 13 clusters in this case, since all data are in stable clusters, and size of the Noise cluster is not very large.

\subsection{Evaluating Classifiers} 
\label{sec:evalclass}
\begin{figure*}[!t]
  \centering
  \includegraphics[width=17cm,height=8cm,keepaspectratio=true,trim={0cm 0cm 0cm 0cm}]{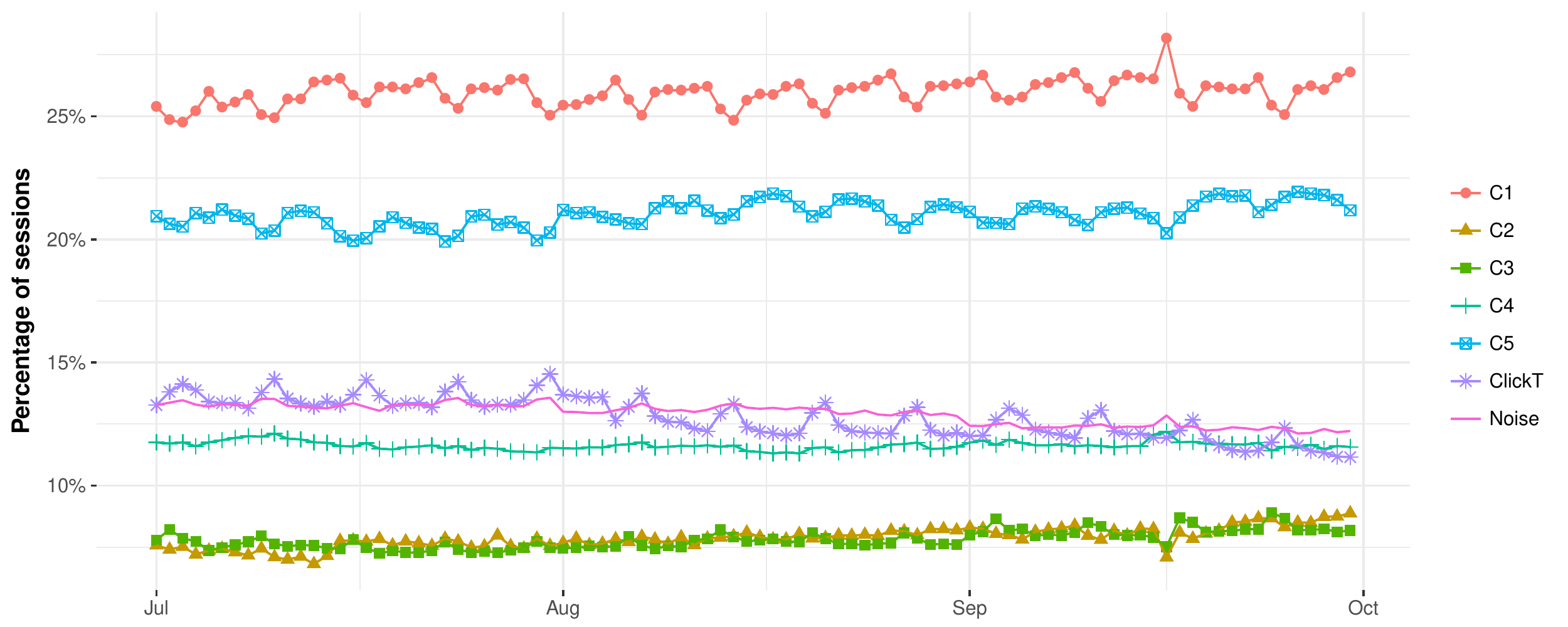}
  \caption{Daily patterns of session categories in a three month period in 2016. Each session category follows a regular weekly seasonal pattern in this period, in spite of hundreds of product experiments on the application. Long term trends are mainly due to annual seasonality. Note that session category names are anonymized for confidentiality reasons, with the exception of Noise and Clickthrough categories.
}
  \label{fig:longtrend}
\end{figure*}
The classifiers assign session categories to each user session. The clusters from previous step are combined into 7 major session categories. This was done for simplification by qualitative assessment of the clusters, as well as to unify the session categories across different Pinterest applications. 

We built each model on 500K sessions on a single day, with \(1/3\) of data assigned as the test set. Table \ref{tab:confusion} shows the confusion matrix for the final classifier for the iPhone application.

Table \ref{tab:classifiererrors} lists the misclassification error of the multiclass random forest classifier in iPhone application for three cases: with all user events, with only Scoring features, and with Scoring features plus the Noise feature (Eq. \ref{eq:ltfeature}). Note that the Noise cluster has a much higher error rate without the new feature. Overall, total classification error in test data was found to be 9.7\% in this device. 

In every device we built a classifier with less than 15\% total classification error. The class with the largest error in every case was the Noise cluster, which is expected, given the features we have removed were the long tail events that were contributing to sessions in this cluster. Given that by definition this class is not very important in our understanding of sessions, the added error in this class is a good trade-off for the added stability.

Figure \ref{fig:longtrend} illustrates the daily proportion of each session category in a three month period in 2016. Most session category names are anonymized for confidentiality reasons, with the exception of Noise and Clickthrough categories. This plot demonstrates stability of the classification results in face of numerous product experiments and user interface changes that are regularly conducted. Additionally, we have the entire data available for over 6 months of activity on Pinterest and have confirmed that the results are robust in time.

\section{Application to Pinterest}
\label{sec:deploymentinsights}

\begin{figure*}[!t]
\begin{subfigure}{5.75cm}
  \centering
  \includegraphics[width=.95\linewidth]{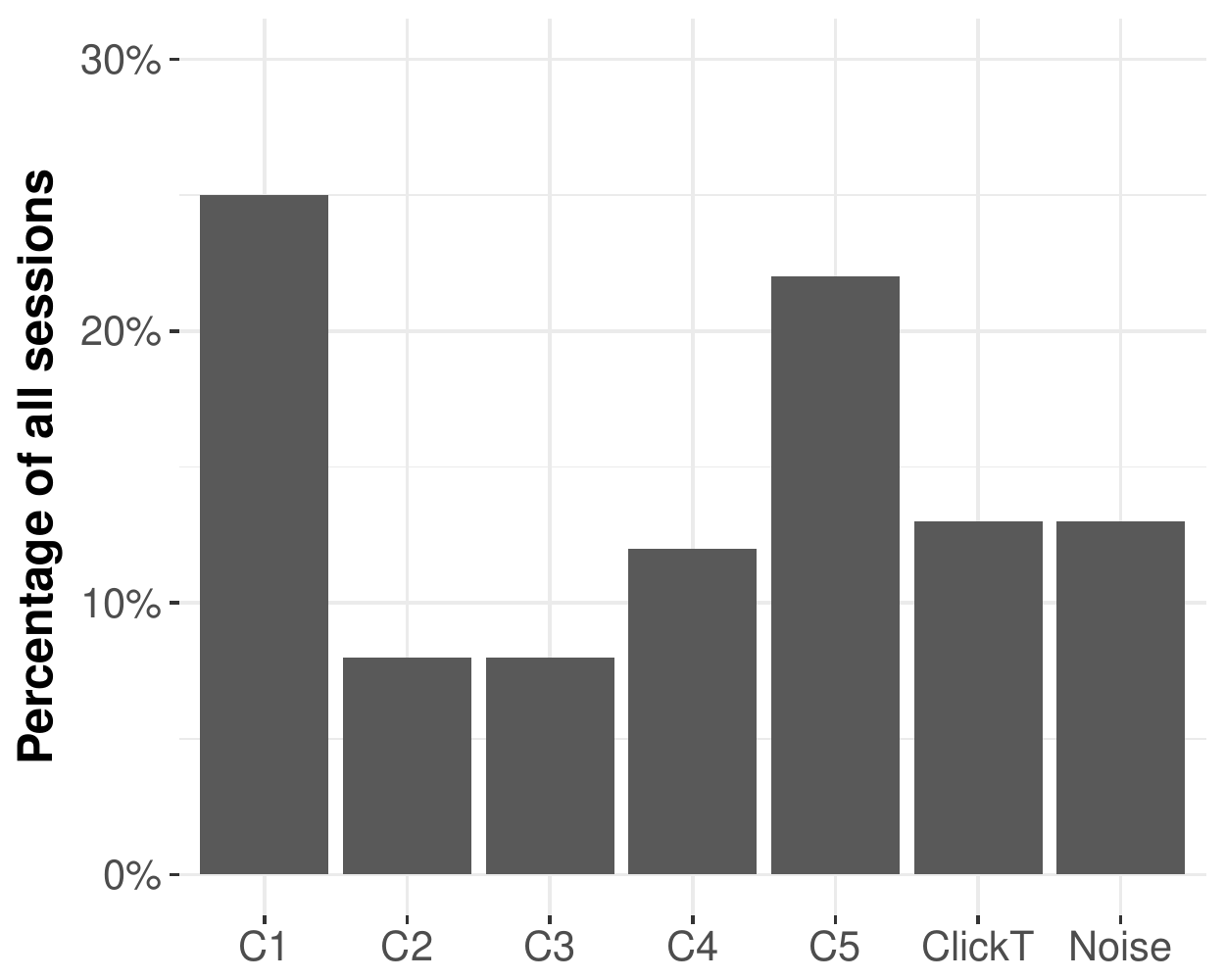}
  \caption{Cluster Sizes}
  \label{fig:sesss}
\end{subfigure}%
\begin{subfigure}{5.75cm}
  \centering
  \includegraphics[width=.95\linewidth]{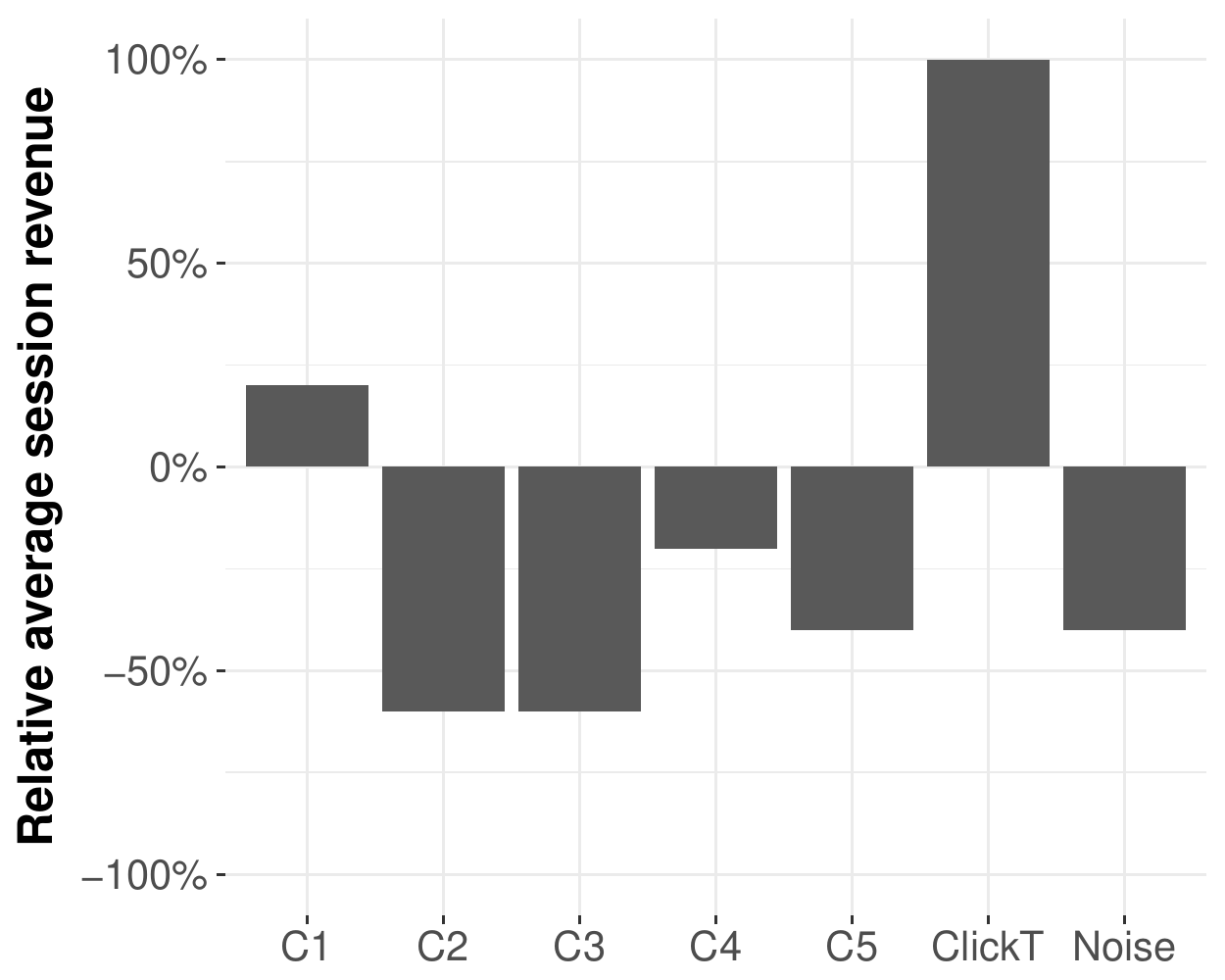}
  \caption{Average Revenue}
  \label{fig:revenue}
\end{subfigure}
\begin{subfigure}{5.75cm}
  \centering
  \includegraphics[width=.95\linewidth]{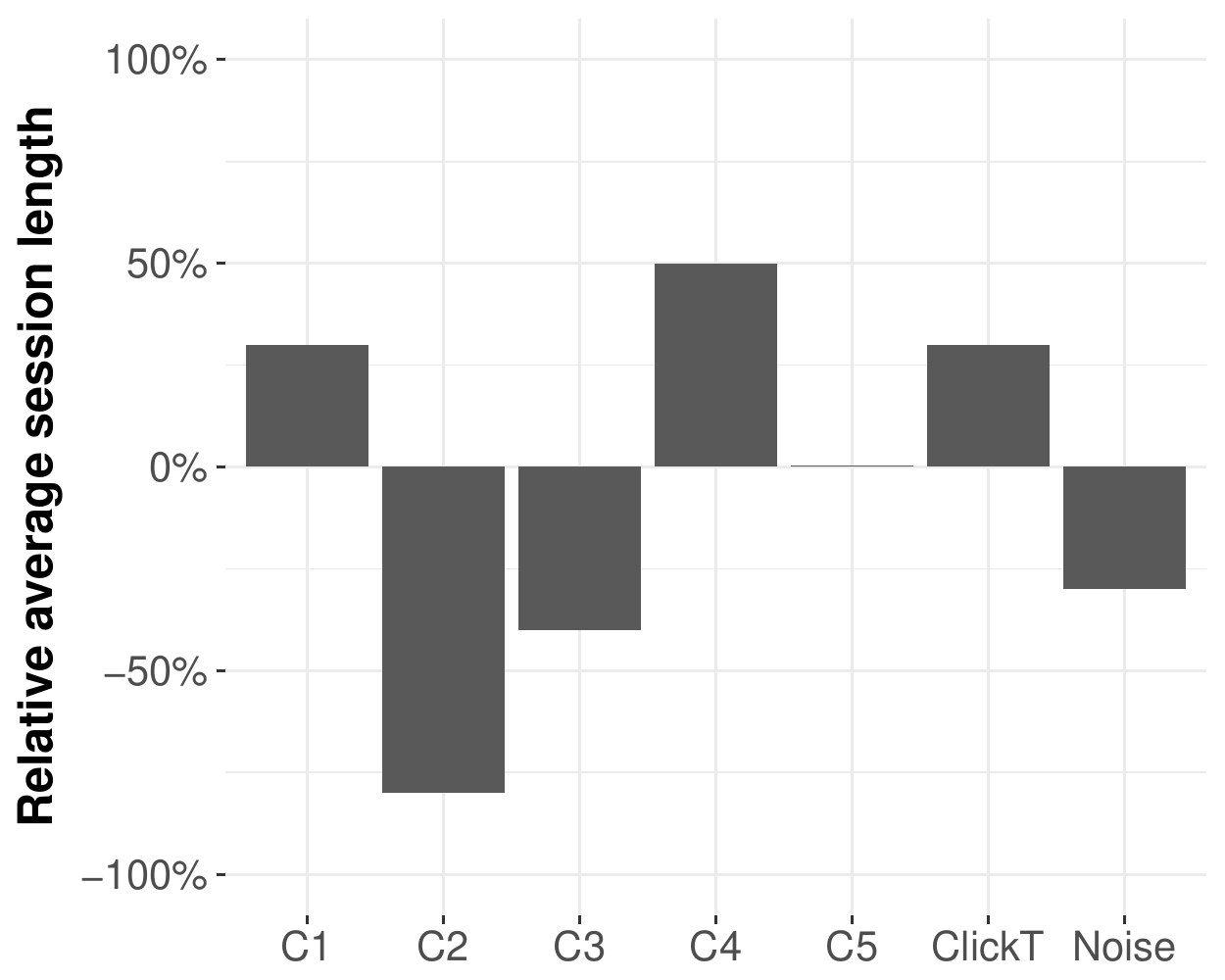}
  \caption{Average Session Length}
  \label{fig:sesslength}
\end{subfigure}
  \caption{(a) Cluster sizes vary from 8\% to 25\%. 13\% of sessions are not classified, i.e. are in \emph{Noise} category. (b) Revenue varies on different session categories. (c) Average session length in each session category, compared to overall average session length. }
  \label{fig:allbar}
\end{figure*}

In this section we present an application of our methodology to Pinterest data \cite{Lo:2016,Cheng:2017}. We created a different set of models for each Pinterest application (iPhone, iPad, Android mobile, Android tablet, web, and mobile Web). In each case, we built models using a sampled set of 500K user sessions on a single day.

For cluster discovery, we used 55 user actions. We found a different number of clusters in each application, e.g. 13 clusters on the iPhone application. After qualitative assessment, we combined the clusters on all applications into 7 major session categories. These session categories are consistent across all devices, and correspond to major session-wise Pinterest use cases. They are named using qualitative assessment of weight of events in each session cluster. The list is as follows: Browse (i.e. mainly viewing content),	Clickthrough (i.e. clicking to the web site that the pin links to),	Notification (i.e. interacting with the in-application notifications),  Retrieval (i.e. viewing content previously saved),	Repin (i.e. saving content), Search (i.e. searching pins), and Noise (i.e. long tail use cases). 

We chose a subset of 12 events as scoring features in the classifiers, as described in Section \ref{sec:classifiers}. The random forest models were built using the randomForest package in R \cite{rf}, with 250 trees and minimum leaf size of 1000 rows. The models were converted to Predictive Model Markup Language (PMML) \cite{guazzelli2009pmml} to be deployed in our distributed pipeline for daily scoring. 

In the following section we present insights found using this work. Due to confidentiality considerations we anonymized session category names in most of the plots, and refer to them by ID instead. The only session categories we will have in every plot are Noise and Clickthrough sessions.

\subsection{Insights} 
The diversity and the number of major use cases of Pinterest was one of the major insights from this work. Here we present some of these insights that were enabled by our session categorization methodology. 

First, Figure \ref{fig:sesss} illustrates the size of each cluster across all devices in one month of data. The smallest clusters are 8\% of sessions, and the biggest ones are more than 20\% of sessions. Note that 13\% of all sessions are long tail forms of engagement, which we labeled Noise session category. 
 
\begin{figure*}[!t]
\begin{subfigure}{9cm}
  \centering
  \includegraphics[width=.9\linewidth]{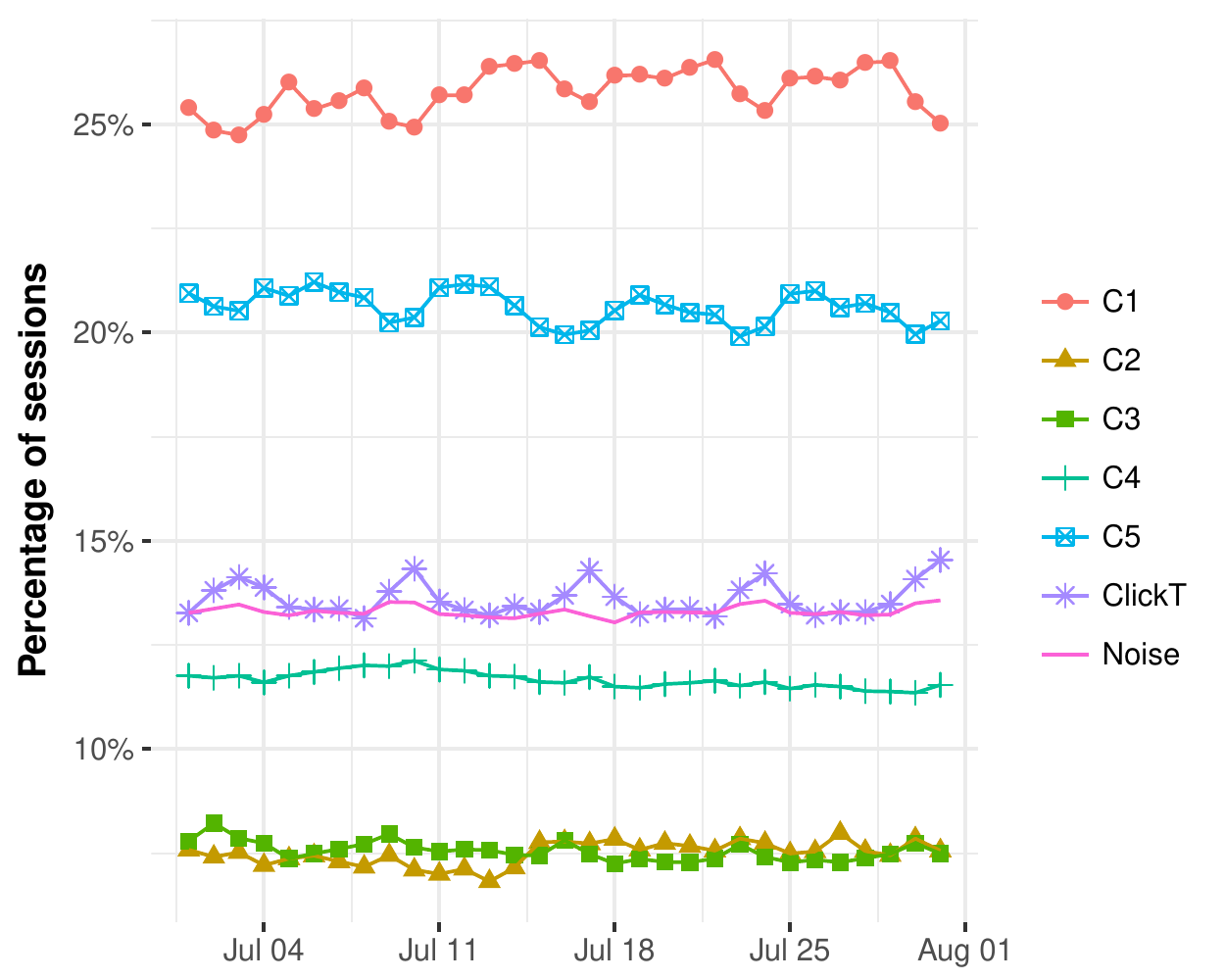}
  \caption{Weekly Patterns}
  \label{fig:trendline}
\end{subfigure}%
\begin{subfigure}{9cm}
  \centering
  \includegraphics[width=.9\linewidth]{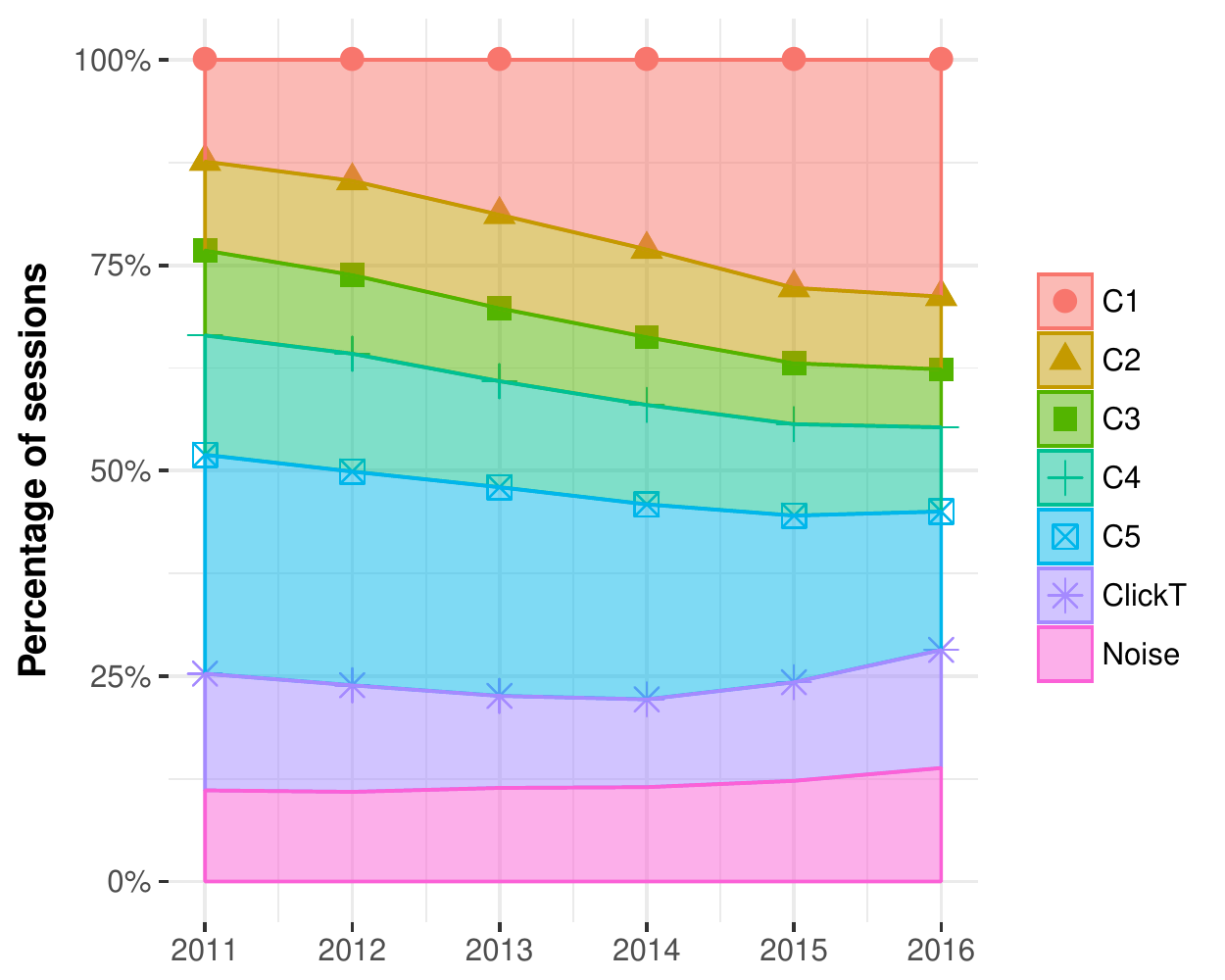}
  \caption{Cohorts}
  \label{fig:cohorts}
\end{subfigure}
  \caption{(a) Proportion of sessions in month of July. C3 and Clickthrough session categories have opposite weekly patterns to C1 and C5, with C3 and Clickthrough peaking on weekends, and C1 and C5 peaking mid-week. (b) Different cohorts have different proportion of session categories, with C5 being used less by newer users and C1 being used a lot more.}
  \label{fig:alltrend}
\end{figure*}

Second, we make an observation in Figure \ref{fig:trendline}, which demonstrates the daily proportion of session categories in the month of July 2016. The proportion of some session categories exhibit weekly seasonality, with Clickthrough and C4 session categories having the opposite weekly pattern to C1 and C5. 

And last, Table \ref{tab:clustweight} lists the highest weighted features in Clickthrough and Search categories. In each case, we list the highest weighted features (i.e. user events) that are associated with each form of engagement. We observe that clusters vary a lot in terms of events that are important in each. For example, Search category has a high average weight for SEARCH\_VIEW event while the Clickthrough category has a high weight for LOAD\_URL event. This demonstrates how the category names were selected.

Next we shall discuss further insights obtained by our clustering methodology in more detail.

\begin{table}[t!]
  \caption{Highest average feature weights in Search and Clickthrough session categories}
  \label{tab:clustweight}
  \begin{tabular}{c c}
    \toprule
  Search Category &	Clickthrough Category  \\
\midrule
	SEARCH\_VIEW, 7.6 & LOAD\_URL, 8.5   \\
	SEARCH\_PINS, 3.0	&BROWSER\_VIEW, 2.8 \\
	VIEW\_END, 1.6	& PIN\_CLICKTHROUGH, 2.4 \\
	DISCOVER\_VIEW, 1.4	& VIEW\_BEGIN , 1.5 \\
   	VIEW\_BEGIN, 1.2	& VIEW\_END, 1.5 \\
	LOAD\_URL, 1.1	&PIN\_VIEW, 0.98 \\
  \bottomrule
\end{tabular}
\end{table}

\xhdr{Cohorts} 
First we examine how Pinterest usage varies among different user cohorts. Figure \ref{fig:cohorts} displays the proportion of each session category by user's cohort, i.e. the year they signed up for Pinterest. We observe that older cohorts and newer cohorts are similar in the proportions of session categories C2 through C4, as well as Clickthrough sessions. However, interestingly session category C1 is utilized much less by older cohorts, and session category C5 is used much more by them. This shows how user behavior has changed on Pinterest as a response to changes in the application as well as cohorts. This finding enables the company to plan for more forward looking product features and functionality.

\xhdr{Revenue varies by session category} 
Next we also examine the amount of revenue generated by sessions of different categories. 
Figure \ref{fig:revenue} illustrates the normalized value of revenue per session, by session category. The values are normalized by average per session revenue. We find significant variation of revenue between different session categories. In particular, Clickthrough sessions and C1 have significantly more revenue per session than others. This finding has important implications in setting company strategy, e.g. in deciding which session-wise use case requires more  investment in product development.

\xhdr{Content}
Another interesting aspect we examine is how the content varies across different session categories. For a specific category of content, namely Food category, we compared pins that were viewed or interacted with on different session categories. We found that pins that were more likely to have been interacted with in Clickthrough sessions are practical recipes and specific ingredients (+10\% more likely, with p \textless 0.05). On the other hand, pins with beautiful pictures, desserts, and aspirational content are more likely to have been engaged with in Repin session category  (difference is +10\%, with p \textless 0.05). 
We observed similar patterns in other content categories as well, where pins with practical content are more likely to be viewed in Clickthrough sessions, and aspirational content in Repin sessions. 

\xhdr{Session length} We also observe that session categories differ significantly by length, which is a metric commonly used to understand depth of engagement. This enables product development to focus on increasing depth of engagement for specific session-wise use cases of the application. This is illustrated in Figure \ref{fig:sesslength}.

\begin{figure}[!t]
  \centering
  \includegraphics[width=8cm,height=8cm,keepaspectratio=true,trim={0cm 4cm 0cm 4cm}]{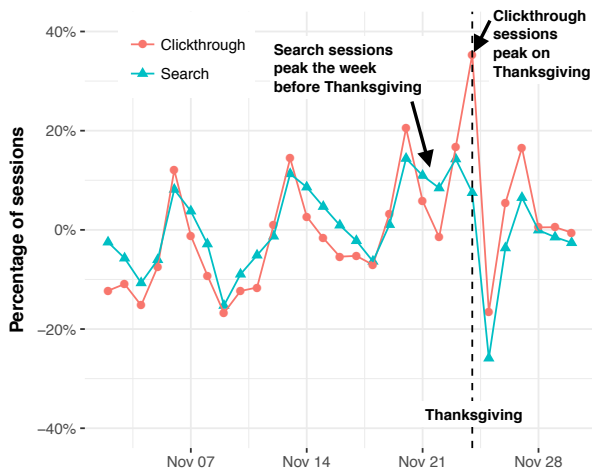}
  \caption{Comparing indexed counts of Search and Clickthrough session categories on and around Thanksgiving holiday in U.S. Count of Search sessions increase a few days prior to the holiday, whereas Clickthrough sessions have a sharp pick on the day.
}
  \label{fig:thanksgiving}
\end{figure}

\xhdr{Thanksgiving daily patterns} Our clustering methodology also enables us to ``zoom-in'' on a particular timeframe to better understand how holidays shape the usage of Pinterest.  In Figure \ref{fig:thanksgiving} we compare the daily counts of the Clickthrough and Search session categories in United States in November 2016, normalized by average daily sessions of each category. The daily patterns around Thanksgiving holiday shows that Clickthrough sessions have a sharp increase on this holiday, which traditionally involves cooking a meal with the extended family. On the other hand, the number of Search sessions increases a few days prior to the holiday, as users start planning for the holiday.

\xhdr{Transition probabilities} Last we examine how users transition between sessions of different categories. The quesiton here is whether users tend to use Pinterest in only a single way or whether there are many different session categories a single user engages in.

\begin{figure}[!t]
  \centering
  \includegraphics[width=7cm,keepaspectratio=true,trim={2cm 1cm 2cm 1cm}]{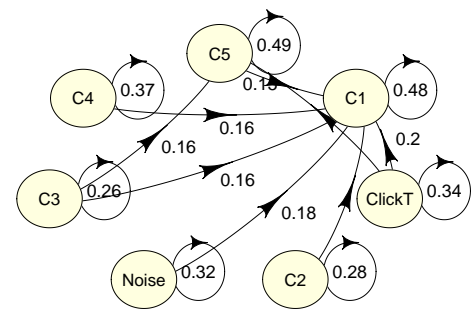}
  \caption{Transition probability of session categories. For simplification the edges with less than 0.15 probability have been removed.}
  \label{fig:transitionp}
\end{figure}

Transition probabilities between session categories are illustrate in Figure \ref{fig:transitionp}. Edges with transition probability less than 0.15 were removed for simplification. Observing the probability of getting the same session category in consecutive sessions, we can see that some session categories are much more likely to be repeated consecutively than others. Additionally, session category C1 is the most likely session to follow all other session categories. 

%\subsection{User Archetypes} Found groups of users who have similar sessions in a period of time. Can show breakdown of users?
%\subsection{Signal for other models} Ads ranking, notification, targeting users in USM. - don't have results for these.

\section{Conclusion}

In this paper we studied the problem of user session categorization, where the goal was to automatically discover categories/classes of user in-session behavior using event logs, and then consistently categorize each user session into the discovered classes. 

We developed a three stage approach which uses clustering to discover categories of sessions, then builds classifiers to classify new sessions into the discovered categories, and finally performs daily classification in a distributed pipeline. An important innovation of our approach was defining a Noise feature that allowed for robust and stable identification of session categories, even though the underlying application is constantly changing. We deployed the solution at Pinterest where it  classifies millions of user sessions daily, and provides actionable insights.

Future work could investigate predicting session categories using only the first few actions which would enable personalized experiences within a single visit. Furthermore, it would be interesting to monitor which user demographic subgroups use the product in a given way, and subsequently connect changes in usage patterns to releases of new product features.

\section{Acknowledgments}
We thank Dan Lurie, Chunyan Wang, and Austin Chang for valuable insights and support throughout the project, Tien Nguyen for help with deployment of the system, and Grace Huang, Dan Frankowski, Brian Karfunkel, Roja Bandari, and Minli Zang for their valuable discussions and insights. 

%Additionally, we thank the brilliant engineers and data scientists at Pinterest for their support of this project. 

%% file: user_session.bbl
\begin{thebibliography}{10}

\bibitem{Kou2012}
Gang Kou and Chunwei Lou.
\newblock Multiple factor hierarchical clustering algorithm forlarge scale web
  page and search engine clickstream data.
\newblock {\em Annals of Operations Research}, 197(1):123--134, 2012.

\bibitem{Katal2013}
A.~Katal, M.~Wazid, and R.~H. Goudar.
\newblock Big data: Issues, challenges, tools and good practices.
\newblock In {\em 2013 Sixth International Conference on Contemporary Computing
  (IC3)}, pages 404--409, Aug 2013.

\bibitem{Mobasher:2000}
Bamshad Mobasher, Robert Cooley, and Jaideep Srivastava.
\newblock Automatic personalization based on web usage mining.
\newblock {\em Commun. ACM}, 43(8):142--151, August 2000.

\bibitem{Eirinaki:2003}
Magdalini Eirinaki and Michalis Vazirgiannis.
\newblock Web mining for web personalization.
\newblock {\em ACM Trans. Internet Technol.}, 3(1):1--27, February 2003.

\bibitem{Yan:1996}
Tak~Woon Yan, Matthew Jacobsen, Hector Garcia-Molina, and Umeshwar Dayal.
\newblock From user access patterns to dynamic hypertext linking.
\newblock In {\em Proceedings of the Fifth International World Wide Web
  Conference on Computer Networks and ISDN Systems}, pages 1007--1014,
  Amsterdam, The Netherlands, The Netherlands, 1996. Elsevier Science
  Publishers B. V.

\bibitem{Cho2002329}
Yoon~Ho Cho, Jae~Kyeong Kim, and Soung~Hie Kim.
\newblock A personalized recommender system based on web usage mining and
  decision tree induction.
\newblock {\em Expert Systems with Applications}, 23(3):329 -- 342, 2002.

\bibitem{srivastava2000web}
Jaideep Srivastava, Robert Cooley, Mukund Deshpande, and Pang-Ning Tan.
\newblock Web usage mining: Discovery and applications of usage patterns from
  web data.
\newblock {\em Acm Sigkdd Explorations Newsletter}, 1(2):12--23, 2000.

\bibitem{silbermann2016}
Ben Silbermann.
\newblock 150 million people finding ideas on pinterest.
\newblock {\em Pinterest Blog}, 2016.

\bibitem{Agrawal:1993}
Rakesh Agrawal, Tomasz Imieli\'{n}ski, and Arun Swami.
\newblock Mining association rules between sets of items in large databases.
\newblock {\em SIGMOD Rec.}, 22(2):207--216, June 1993.

\bibitem{Cooley}
R.~Cooley, B.~Mobasher, and J.~Srivastava.
\newblock Web mining: information and pattern discovery on the world wide web.
\newblock In {\em Proceedings Ninth IEEE International Conference on Tools with
  Artificial Intelligence}, pages 558--567, Nov 1997.

\bibitem{Tiwari}
S.~Tiwari, D.~Razdan, P.~Richariya, and S.~Tomar.
\newblock A web usage mining framework for business intelligence.
\newblock In {\em 2011 IEEE 3rd International Conference on Communication
  Software and Networks}, pages 731--734, May 2011.

\bibitem{Liu}
Jinze Liu, S.~Paulsen, Wei Wang, A.~Nobel, and J.~Prins.
\newblock Mining approximate frequent itemsets from noisy data.
\newblock In {\em Fifth IEEE International Conference on Data Mining
  (ICDM'05)}, pages 4 pp.--, Nov 2005.

\bibitem{Shaw}
Michael~J. Shaw, Chandrasekar Subramaniam, Gek~Woo Tan, and Michael~E. Welge.
\newblock Knowledge management and data mining for marketing.
\newblock {\em Decis. Support Syst.}, 31(1):127--137, May 2001.

\bibitem{Mobasher}
Bamshad Mobasher, Honghua Dai, Tao Luo, and Miki Nakagawa.
\newblock Effective personalization based on association rule discovery from
  web usage data.
\newblock In {\em Proceedings of the 3rd International Workshop on Web
  Information and Data Management}, WIDM '01, pages 9--15, New York, NY, USA,
  2001. ACM.

\bibitem{Agrawal:1995}
Rakesh Agrawal and Ramakrishnan Srikant.
\newblock Mining sequential patterns.
\newblock In {\em Proceedings of the Eleventh International Conference on Data
  Engineering}, ICDE '95, pages 3--14, Washington, DC, USA, 1995. IEEE Computer
  Society.

\bibitem{Shahabi}
C.~Shahabi, A.~M. Zarkesh, J.~Adibi, and V.~Shah.
\newblock Knowledge discovery from users web-page navigation.
\newblock In {\em Proceedings of the 7th International Workshop on Research
  Issues in Data Engineering (RIDE '97) High Performance Database Management
  for Large-Scale Applications}, RIDE '97, pages 20--, Washington, DC, USA,
  1997. IEEE Computer Society.

\bibitem{Han:1992}
Jiawei Han, Yandong Cai, and Nick Cercone.
\newblock Knowledge discovery in databases: An attribute-oriented approach.
\newblock In {\em Proceedings of the 18th International Conference on Very
  Large Data Bases}, VLDB '92, pages 547--559, San Francisco, CA, USA, 1992.
  Morgan Kaufmann Publishers Inc.

\bibitem{Fu}
Yongjian Fu, Kanwalpreet Sandhu, and Ming-Yi Shih.
\newblock A generalization-based approach to clustering of web usage sessions.
\newblock In {\em Revised Papers from the International Workshop on Web Usage
  Analysis and User Profiling}, WEBKDD '99, pages 21--38, London, UK, UK, 2000.
  Springer-Verlag.

\bibitem{Salton:1986}
Gerard Salton and Michael~J. McGill.
\newblock {\em Introduction to Modern Information Retrieval}.
\newblock McGraw-Hill, Inc., New York, NY, USA, 1986.

\bibitem{Heer02miningthe}
Jeffrey Heer, Ed~H. Chi, and H.~Chi.
\newblock Mining the structure of user activity using cluster stability.
\newblock In {\em in Proceedings of the Workshop on Web Analytics, SIAM
  Conference on Data Mining (Arlington VA}. ACM Press, 2002.

\bibitem{Jin:2004}
Xin Jin, Yanzan Zhou, and Bamshad Mobasher.
\newblock Web usage mining based on probabilistic latent semantic analysis.
\newblock In {\em Proceedings of the Tenth ACM SIGKDD International Conference
  on Knowledge Discovery and Data Mining}, KDD '04, pages 197--205, New York,
  NY, USA, 2004. ACM.

\bibitem{plsa}
Guandong Xu, Yanchun Zhang, and Xiaofang Zhou.
\newblock Using probabilistic latent semantic analysis for web page grouping.
\newblock In {\em 15th International Workshop on Research Issues in Data
  Engineering: Stream Data Mining and Applications (RIDE-SDMA'05)}, pages
  29--36, April 2005.

\bibitem{Spiliopoulou:2003}
Myra Spiliopoulou, Bamshad Mobasher, Bettina Berendt, and Miki Nakagawa.
\newblock A framework for the evaluation of session reconstruction heuristics
  in web-usage analysis.
\newblock {\em INFORMS J. on Computing}, 15(2):171--190, April 2003.

\bibitem{kaufman1987clustering}
Leonard Kaufman and Peter Rousseeuw.
\newblock {\em Clustering by means of medoids}.
\newblock North-Holland, 1987.

\bibitem{Rousseeuw:1987:SGA:38768.38772}
Peter Rousseeuw.
\newblock Silhouettes: A graphical aid to the interpretation and validation of
  cluster analysis.
\newblock {\em J. Comput. Appl. Math.}, 20(1):53--65, November 1987.

\bibitem{Lange}
Tilman Lange, Volker Roth, Mikio~L. Braun, and Joachim~M. Buhmann.
\newblock Stability-based validation of clustering solutions.
\newblock {\em Neural Computation}, 16(6):1299--1323, 2017/01/29 2004.

\bibitem{Hennig07cluster-wiseassessment}
Christian Hennig.
\newblock Cluster-wise assessment of cluster stability.
\newblock {\em Computational Statistics and Data Analysis}, pages 258--271,
  2007.

\bibitem{Lo:2016}
Caroline Lo, Dan Frankowski, and Jure Leskovec.
\newblock Understanding behaviors that lead to purchasing: A case study of
  pinterest.
\newblock In {\em Proceedings of the 22Nd ACM SIGKDD International Conference
  on Knowledge Discovery and Data Mining}, KDD '16, pages 531--540, New York,
  NY, USA, 2016. ACM.

\bibitem{Cheng:2017}
Justin Cheng, Caroline Lo, and Jure Leskovec.
\newblock Understanding behaviors that lead to purchasing: A case study of
  pinterest.
\newblock In {\em Proceedings of the 26th International Conference on World
  Wide Web}. International World Wide Web Conferences Steering Committee, 2017.

\bibitem{rf}
Andy Liaw and Matthew Wiener.
\newblock Classification and regression by randomforest.
\newblock {\em R News}, 2(3):18--22, 2002.

\bibitem{guazzelli2009pmml}
Alex Guazzelli, Michael Zeller, Wen-Ching Lin, Graham Williams, et~al.
\newblock Pmml: An open standard for sharing models.
\newblock {\em The R Journal}, 2009.

\end{thebibliography}
